\documentclass[conference]{IEEEtran}
\usepackage{amsmath}
\usepackage{algorithmic}
\usepackage{array}
\usepackage{grffile}
\usepackage{epstopdf}
\usepackage{wrapfig}
\usepackage{color,soul}

\usepackage{stfloats}
\usepackage{url}
\begin{document}
\title{Outlier Detection in Smart Grid Communication}

\author{\IEEEauthorblockN{Nelson Makau Mutua}
\IEEEauthorblockA{Faculty of Information Technology\\
Brno University of Technology\\
Brno, Czech Republic\\
Email: imutua@fit.vut.cz}
\and
\IEEEauthorblockN{Petr Matou\v{s}ek}
\IEEEauthorblockA{Faculty of Information Technology\\
Brno University of Technology\\
Brno, Czech Republic\\
Email: matousp@fit.vutbr.cz}}


\maketitle

\begin{abstract}
Industrial Control System (ICS) networks transmit control and monitoring data in critical environments such as smart grid. 
% A smart grid network is significant for proper functioning of critical infrastructures for instance power plant, water treatment facilities and gap pipeline. 
Cyber attacks on smart grid communication may cause fatal consequences on energy production, distribution, and eventually the lives of people. 
%In  spite  of  the  fact that  security  of  such  systems deserves  attention,  application  of  thorough  security  intelligence approaches  to smart grid  is  not  a  standard  practice. 
Since the attacks can be initiated from both the inside and outside of the network, traditional smart grid security tools like firewalls or Intrusion Detection Systems (IDS), which are typically deployed on the edge of the network, are not able to detect internal threats. For this reason, we also need to analyze behavior of internal ICS communication. 
% where they  inspect incoming and outgoing traffic. This strategy is adequate for dealing with external threats. In the case of internal threats caused, for example, by an infected control station, it is difficult to detect malicious activity at the network edge. 
%, which is frequently disguised as legitimate communication. 

Due to its nature, ICS traffic exhibits stable and predictable communication patterns.  These patterns can be described using statistical models. By observing selected features of ICS network communication like packet inter arrival times, we can create a statistical profile of the communication based on the patterns observed in the normal communication traffic. This technique is effective, fast and easy to implement. As our experiments show, statistical-based anomaly detection is able to detect common security incidents in ICS communication. This paper employs selected network packet attributes to create a statistical model for anomaly detection using the Local Outlier Factor (LOF) algorithm. The proof-of-concept is demonstrated on IEC 60870-5-104 (a.k.a.  IEC 104) protocol.

\end{abstract}

% \fi
\IEEEpeerreviewmaketitle

\section{Introduction}
ICS communication provides proper functioning of critical infrastructure systems. These systems are naturally exposed to external threats including cyber attacks \cite{leith-13-identification}. 
% public networks that are at a high risk of cyber-attack. The control procedures of critical infrastructures such as power plant, oil and gas facilities, chemical processing plants, traffic control systems, among others are becoming more and more vulnerable to external network threat 
% Being less protected and more sensitive to software updates than a typical office environment makes additional security measures smart grid necessary. 
% Also, ease of exposure to cyber-attacks once the physical levels of security is breached (e.g. insider attacks) requires a new look at how to protect these critical systems. 
Traditionally, industrial systems are well protected against external threats through the use of firewalls and IDS devices that filter communication between ICS systems and Internet traffic, making direct attacks on smart grid communication a rare case. However, attackers can gain access to the system by sending malware to a user via an infected e-mail attachment \cite{lee-16-analysis}.

The importance of the research is motivated by recent cyber attacks against critical infrastructure systems. One of the attacks against Ukrainian power company happened in December 23, 2015 when BlackEnergy malware caused power disruption to 225,000 customers, lasting up to 6 hours \cite{lee-16-analysis}. This  attack included multiple stages starting with “spear phishing” e-mails targeting a staff to gain access to the corporate network of the power company. Once inside the power company network, attackers gathered credentials and used VPNs to get access to the internal network. More recent ransomware attack against the Colonial Gas Pipeline in the U.S. happened in May 2021 \cite{tidy_2021}. Similarly to BlackEnergy attack, it was also initiated from an infected internal station.

Compared to standard information and communication systems, smart grid communication exhibits stable, periodical, and regular communication patterns since the communication occurs between devices with no or little human interference. Typically, a controlling station periodically requests status data from a field device like the Programmable Logical Controller (PLC) or Remote Terminal Unit (RTU) in order to provide a real-time view on industrial processes. 

To detect internal threats, we need (i) to regularly monitor smart grid communication and (ii) observe suspicious patterns that occur in the network traffic. One solution is to employ ICS monitoring using extended IPFIX protocol\cite{matousek-19-increasing} that retrieves monitoring data about active ICS communication. To detect unknown adverse events or unusual behavior, we can observe statistical patterns using ICS flow data. 

% This makes anomaly detection based on statistical modelling of IEC 104 packets inter-arrival time a potentially viable approach to detect unknown adverse events that might be caused by insider threats.

%Unlike our previous research, \cite{matouvsek2019increasing} where we modeled sequences of ICS messages using probabilistic automata, in this paper, we focus on the timing properties of ICS communication. 
In this research, we closely examine statistical distribution of inter-arrival times of IEC 104 packets. This is a part of my PhD study that is focused on anomaly detection of ICS communication using statistical methods. The main idea behind the research comes out from  stable communication patterns that are typical for ICS communication and can be observed on packet and flow levels. Within the PhD. research, we plan to apply various statistical methods on smart grid communication and evaluate their efficiency for covering common ICS attack vectors defined by MITRE ATT\&CK for ICS matrix\footnote{See \url{https://collaborate.mitre.org/attackics/index.php/Main_Page} [06/2021]}. 

\subsection{Contribution}
This research paper presents a proof-of-concept to outlier detection in smart grid communication. We observe packet inter-arrival times of IEC 104 communication and  create a statistical profile of the normal traffic. Then we apply the Local Outlier Factor (LOF) algorithm to detect outliers which represent anomalies. Our preliminary results proves viability of this approach for statistical-based anomaly detection in ICS communication. 

%and future directions of the research. 

%This technique was able to detect outliers in a smart grid network communication as demonstrated in section IV.

% We created the tool as a prototype in python programming language using LOF algorithm. Although our key goal was to test the applicability of our proposed solution, we also tested the performance of the developed tool. To get an idea of how our tool is, we gathered some basic time statistics. The processing time of extracting the key attributes from the PCAP file into a CSV file took an average of one minute and 10 seconds. This time had small deviations because it takes time to initialize Tshark. The amount of time required to learn a CSV file depends on the number of packets. At average, to learn 5,000 packets, 20 seconds were required. The proposed outlier detection method is implemented as a part of the anomaly detection module of the IPFIX monitoring system for ICS domain.
% \enlargethispage{4mm}

\subsection{Structure of the Paper}
This paper is structured as follows. In Section \ref{sec:related-work} we give an overview of published works related to statistical anomaly detection in ICS networks. Next, we briefly introduce IEC 104 protocol, statistical features and the LOF algorithm. Section \ref{sec:preliminary-results} shows our preliminary results. The last section concludes the work and discusses future steps. 
%Further we describe dataset and the techniques of detecting outliers in smart grid communication in section III. Then we describe the preliminary results of our research in section IV, and finally, we conclude the paper in section V. 

\section{Related Work}\label{sec:related-work}
Exploitation of timing attributes such as average packet inter-arrival times and the number of packets or bytes transmitted in a certain interval for anomaly detection has been studied in the past. Barbosa et al. \cite{barbosa-12-first,barbosa-14-anomaly} investigated the use of spectral analysis to uncover traffic periodicity. Udd et al. \cite{udd2016exploiting} examined a TCP sequence prediction attack for the IEC 104 protocol. Other works mostly focused on flooding and DoS attacks \cite{bhatia2014practical}. We explore what attack vectors described by MITRE ATT\&CK for ICS can be covered by LOF method.
% The main limitation of these approaches is the “semantic gap”. These approaches demonstrate that traffic patterns can be used for anomaly detection but provide little insights about which packets caused the anomalies.

To enhance detection ability of cyber attack, deep packet inspection (DPI) implemented in IDS tools and timing models have been proposed. Sayegh et al. \cite{sayegh2014scada} modelled the inter-arrival times between signatures (i.e., packet sequences) and validated this technique with large amount of injected signatures. Barbosa et al. \cite{barbosa2016exploiting} proposed to model the period of repeated requests in an orderless group. They evaluated the approach on Modbus and MMS datasets without attacks and set relaxed thresholds to avoid high false positive rates. This prevents the detection of subtle changes within a single period which is covered by our approach.

More recently, sequence-aware approaches have been explored. Yang et al. \cite{yang2014stateful}, Goldenberg and Wool \cite{goldenberg2013accurate}, and Kleinmann and Wool \cite{kleinmann2016automatic} used finite automata to model message sequences of IEC 104, Modbus TCP and S7 respectively. Casselli et al. \cite{caselli2015sequence} modelled a sequence of messages using discrete-time Markov chains in order to detect sequence attacks. These approaches observe the order of messages. Since they require deep packet analysis, they have high demands on processing. Modeling of statistical behavior as proposed in this paper is fast and simple for implementation while giving comparable results.

\section{Anomaly Detection in Smart Grid Networks}\label{sec:anomaly-detection}
\subsection{Smart Grid Communication}\label{sec:smart-grid-communication}
Industrial protocols used in smart grid communication include protocol IEC 61850 (GOOSE, MMS), Modbus, IEC 104, DNP3, DLMS, and others \cite{knapp2014industrial}. These protocols transmit control and status data from industrial processes running on RTUs or IEDs. Protocols like IEC 104, DNP3, MMS or DLSM communicate using a {\em client-server model}. A master (controlling) station sends commands to a RTU slave (controlled station) in control direction while the slave delivers monitoring data in monitor direction. 
% (RTU monitors or commands a controlled station (RTU slave) in this model.ICS client-server communication model is delivered in the monitoring direction i.e. from the controlled station (slave) to the controlling station (master) or in the control direction i.e. controlling station (master) to the controlled station (slave). 
Protocols like GOOSE or Modbus use a {\em publish-subscribe mechanism} in which an application writes data into a local buffer, which is then periodically transmitted to a subscribed agent via L2 multicast. % The sub-sections below briefly review the main features of IEC 104 and GOOSE protocols. Also, the inter-arrival time used in our statistical model is defined and our dataset discussed.

\subsubsection{IEC 104}
%\vspace{5mm} %5mm vertical space
% This SCADA protocol is defined within a collection of standards called IEC 60870. This protocol is a TCP/IP adaptation of the long-standing IEC 101 serial protocol, also known as IEC60870-5-101. IEC 101 defines the remote-control functionalities needed in extensive areas. Within the electrical industry, both IEC 104 and IEC 101 are used extensively to help in establishing communications links that connect the electrical control stations to the substations \cite{kerkers-18-tool}. 

IEC 104 protocol is an application protocol that consists of Application Protocol Control Information (APCI) and Application Service Data Unit (ASDU) sub-layers. It is implemented over TCP but for monitoring purposes we observe so-called virtual flows that represent records with a single ASDU packet transmitted on wire. We focus on inter-arrival times but there are additional attributes that can be used for statistical model: ASDU size which is also stable for specific IEC 104 commands, frequency of selected commands, e.g., spontaneous events, activations, packet size, etc.

\subsubsection{Packets inter-arrival time}
\begin{figure}[h]
    \centering
    \includegraphics[width=5cm]{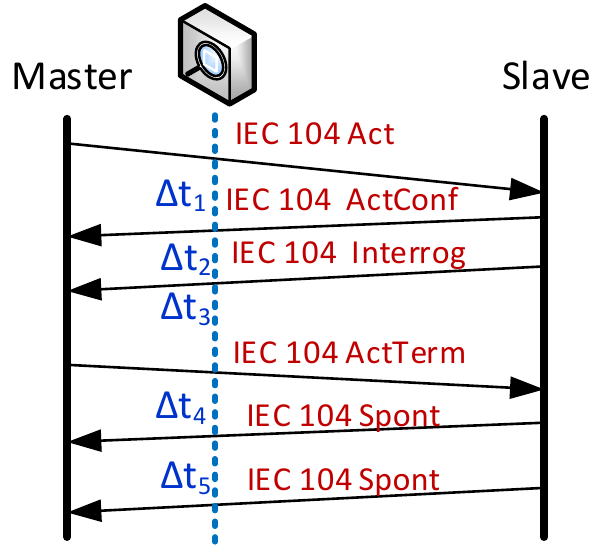}
    \caption{Observing inter-arrival times in IEC 104 communication.}
    \label{fig:inter-arrival-times}
\end{figure}

Packets inter-arrival time is the time taken between the arrival of two subsequent packets. It is computed as a difference between timestamps of two these packets. Inter-arrival time is characteristic for ICS transmissions and exhibits stable and predictable patterns. Changes in inter-arrival times indicate an anomaly on the communication channel. As demonstrated in our previous research \cite{FITPUB12385}, additional features like packet size or modeling of exact IEC 104 commands do not improve the accuracy of the statistical approach. We plan to confirm this observation by future experiments. For modelling inter-arrival time of industrial protocols with master-slave communication profile, it is natural to observe statistical distribution of bi-directional transmissions as depicted in Fig.  \ref{fig:inter-arrival-times}.
%every industrial communication, it is possible to model the distribution of inter-arrival times in bi-directional transmissions, see Fig. \ref{fig:inter-arrival-times}. This depends on the underlying behavior of the ICS protocol, e.g. bi-directional distribution gives the sense for master-slave communication.   
%\vspace{-5mm}

%\vspace{-2mm}
\subsubsection{Dataset}
In our experiments, we work with the IEC 104 protocol that is commonly used in smart grids for substation control. We utilise four datasets obtained from the Brno University of Technology testbed, see Tab. \ref{tab:datasets}. In order to test the ability of attack detection we created a set of ICS flow records with emulated attack and a AD testing tool \footnote{The tool and dataset are available at \url{https://github.com/nelsonmakau/anomaly-detection-in-smart-grid-communication}.}. 

\begin{table}[h]
\centering
\begin{tabular}{ |l||c|c|c|}
 \hline
 Dataset & Packets & Duration & Devices \\
 \hline
 10122018-104Mega & 104,534   & 4h 53min & 4 \\
 13122018-mega104 &  1,460,829  & 71h 17min & 14\\
 mega104-14-12-18 & 14,597 & 15h 38min &  2 \\
 mega104-17-12-18 & 58,931 & 67h 55min &  2 \\
 \hline
\end{tabular} 
  \caption{IEC 104 datasets for experiments.}\label{tab:datasets}
  \vspace{-4mm}
\end{table}

%\vspace{-6mm}
\subsection{Statistical Outlier Detection}\label{sec:outlier-detection}
Outlier detection is a statistical technique that discovers data points that are inconsistent with the rest of the data. In general, outliers and inliers are determined using a data distribution model. This paper focuses on unsupervised  outlier detection. 

%The following are two key types of statistical outlier detection techniques: supervised and unsupervised. 
% Supervised algorithms assume that the data contains fully labeled training and test datasets. A significant approach in this situation is to create a predictive model for non-normal and normal data classes. Every hidden data point is compared to the model to determine which class it belongs to. The main issue of supervised outlier detection is that the outlier data points are much fewer than normal data points in training datasets. Support Vector Machine (SVM) is a popular algorithm used in the supervised learning. 
Unsupervised methods do not require data labels. As a result, they are more adaptable. The fundamental idea behind unsupervised outlier detection is to score data points solely on the essential characteristics of the dataset. In general, density or distance are used to determine whether a data point is an inlier (normal) or outlier (anomaly). In this proof-of-concept study, we apply Local Outlier Factor (LOF) method \cite{breunig-00-lof}.

%LOF is a popular algorithm used in the unsupervised learning. 
The scientific research on statistical outlier detection provides two approaches to handle outliers in a dataset. First, outliers must be identified for further investigative process. Second, the data model should be designed to handle outlier data points accurately. % In this research, the LOF algorithm was used for outlier detection.

\subsection{Detecting Outliers in Smart Grid}\label{sec:detecting-outliers}
Outlier detection is a statistical procedure that finds suspicious events or items that differ from a dataset's normal form. Outliers are detected as data points that have a significantly lower density than their neighbors. 
%It has sparked a lot of interest in the fields of data mining and machine learning. 
The purpose of outlier detection is to detect rare events or unusual activities that differ from the majority of data points in a dataset \cite{boukerche2020outlier}. %Outlier detection has recently emerged as a critical issue in a variety of applications, including health care, credit card fraud detection, and intrusion detection in computer networks. 
Outlier detection can detect global or local outliers. For a global outlier, outlier detection considers all data points, and the data point is considered an outlier if it is far away from all other data points. The local outlier detection covers a small subset of data points at a time. A local outlier is based on the probability of data point being an outlier as compared to its local neighborhood which is measured by the k-Nearest Neighbors (kNN) algorithm. LOF is a density-based unsupervised anomaly detection method that computes a given data point's local density deviation with respect to its neighbors. LOF scores are computed for all data points according to parameter $k$ (the number of nearest neighbors) as follows \cite{alghushairy-21-review}: 

%\vspace{5mm} %5mm vertical space
\textbf{Definition 1:} \emph{d(p, o) is the Euclidean distance between two data points p and o.} 
The distance between two data points $p$ and $o$ is calculated using an Euclidean n-dimensional space: %, see Equation \eqref{eq:distance}. 
\begin{eqnarray}
  d(p,o) & = & \sqrt{\sum_{i=1}^n (p_i - o_i)^2} \label{eq:distance}
\end{eqnarray}
Let $D$ be a dataset and $k$ a positive integer. For a data point $p$, the \emph {k-distance (p)} is the distance \emph {d(p,o)} between $p$ and the farthest neighbor data point $o$ by the following conditions:
\begin{enumerate}
    \item At the least, $k$ data points (records) $o' \in D \backslash \{p\}$ maintains that $d(p,o') \le d(p,o)$.
    \item At the most, $k-1$ data points (records) $o' \in D \backslash \{p\}$ maintains that $d(p,o') < d(p,o)$.
\end{enumerate}

\textbf{Definition 2:} \emph{k-Nearest Neighbors of p}. The meaning of k-Nearest Neighbors of $p$ is any data point $q$ whose distance to the $p$ data point is not greater than the \emph{k-distance(p)}.Those k-nearest neighbors of $q$ form the so called {\em k-distance neighborhood of p}, as described below:
\begin{eqnarray}
N_{k-distance(p)} (p) = \{q\epsilon D/\{p\}|d(p,q)\leq k-distance(p)\} 
\end{eqnarray}

\textbf{Definition 3:} \emph {Reachability distance of $p$ with respect to $o$.} Let $k$ be a positive integer. The reachability distance of data point $p$ with regard to $o$ is as follows:
\begin{eqnarray}
reach-dist_k(p,o) = max \{k-distance(o),d(p,o)\} \label{eq3:reachability}
\end{eqnarray}

\begin{figure}[ht]
    \centering
    \includegraphics[width=7cm]{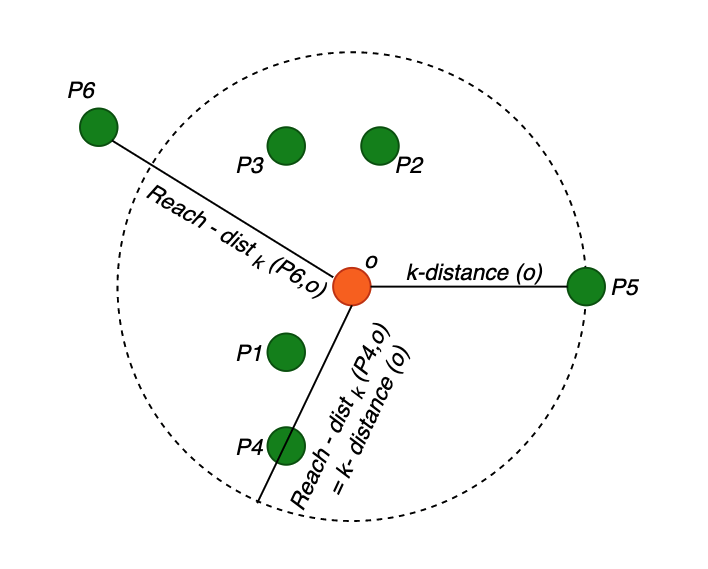}
    \caption{Reachability distance for different data points }
    \label{fig:distance}
\end{figure}

The principle of LOF reachability is depicted in Fig. \ref{fig:distance}, further details are available at \cite{breunig-00-lof}.

\vspace{-2mm}

\vspace{-2mm}
\section{Preliminary results}\label{sec:preliminary-results}
\begin{figure}[h]
    \centering
    \includegraphics[width=9cm]{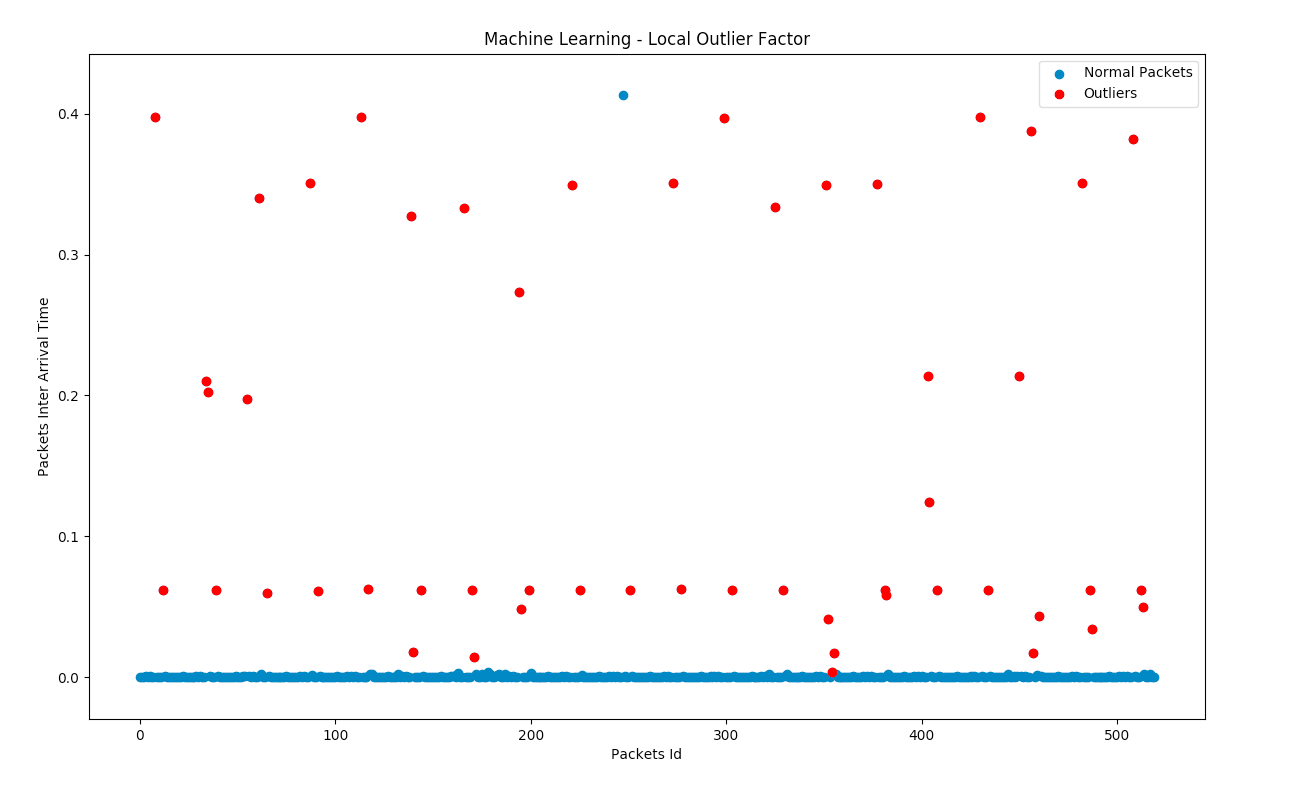}
   \caption{Anomaly detection in 10122018-104Mega PCAP file}
    \label{fig:anomaly-detection1}
    %\vspace{-2mm}
\end{figure}

\begin{figure}[h]
    \centering
    \includegraphics[width=9cm]{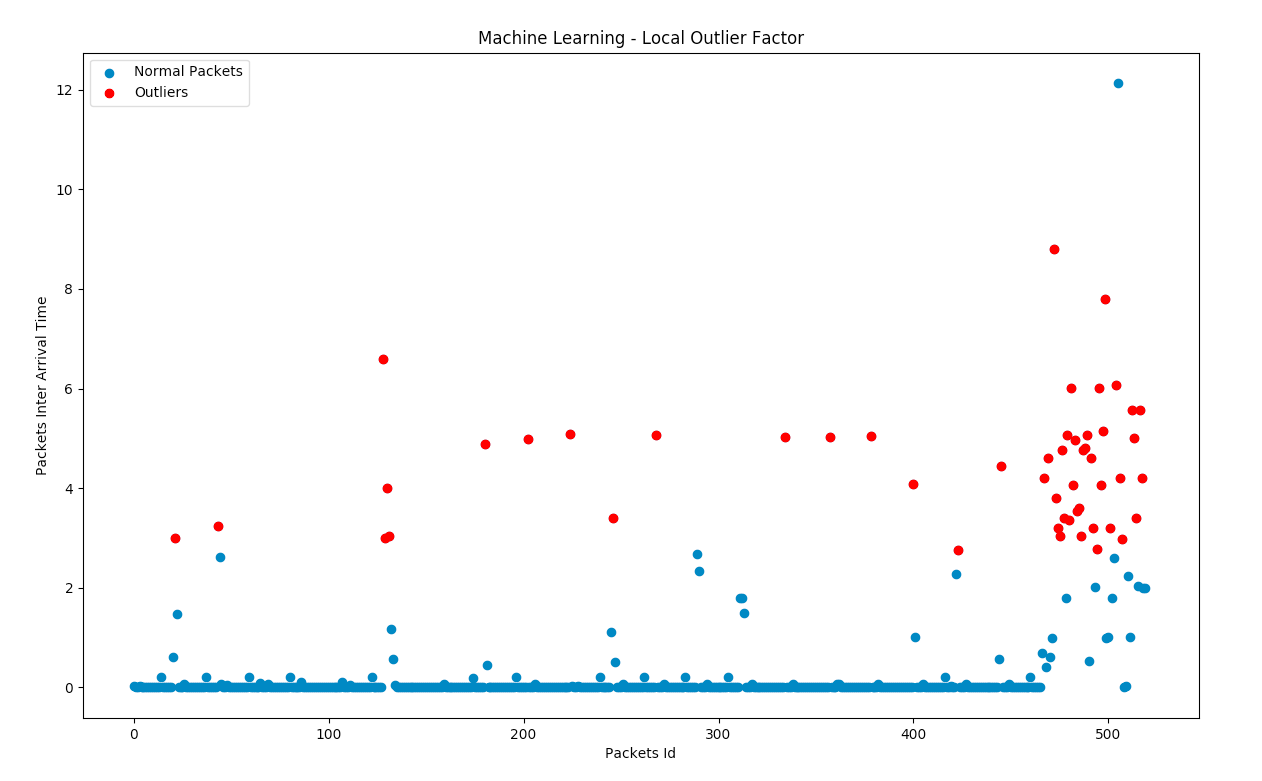}
    \caption{Anomaly detection in mega104-14-12-18 PCAP file}
    \label{fig:anomaly-detection2}
    %\vspace{-2mm}
\end{figure}

We created a prototype\footnotemark[2] tool implementing LOF algorithm. We applied LOF on IEC 104 inter-arrival times in order to check if LOF modeling is suitable for ICS communication. 
% how suitable this statistical method  is on statistical attributes of ICS communication. 
%the suitability of our detection method against outliers using the dataset described in section III-A4. 
We used LOF algorithm to learn inter-arrival time distribution of the transmitted IEC 104 packets by computing the k-distance, reachability distance and density of data points. The algorithm uses the learned model to raise an alarm for packets that  significantly differ from the model. Figures \ref{fig:anomaly-detection1} and \ref{fig:anomaly-detection2} present a graphical representation of inter-arrival times of two datasets. The red points denote inter-arrival time distribution detected as anomaly (outliers) while the blue points represent normal communication. 

We discovered that applying LOF to a large data stream is extremely computationally inefficient and may lead to incorrect prediction results. We then applied the LOF on  time windows where each data block contains 5000 inter-arrival time records.

The main advantage of the LOF algorithm is that it works well with stream data and detects outliers with respect to density of their neighbouring data records. Also, this algorithm is able to detect outliers regardless of the data distribution of normal behavior, since it does not make any assumptions about the distribution of data records. 
%A statistical model is sensitive to outliers which are particular data with exceptionally low probability that may be incorrectly marked as outliers. 
Our preliminary results show that the proposed method works well with ICS data and is easy to operate. 
% can detect outliers in network communication and it is easier to operate.

Validation tests were applied to determine the viability of the proposed technique. We tested the stability of the approach using our datasets. For our validation tests, we divided normal communication into two parts. We used the first two thirds of the communication to train the model. Then, we tested these ranges on the last third of data. After testing, we classified all testing packets as normal communication.  The results demonstrate high accuracy of the proposed approach. The method did not produce any false positive during the test.

\section{Conclusion}\label{sec:conclusion}
%In the recent past, industrial systems have become a key target for attackers. Therefore, good security mechanisms are paramount in these systems.
In this paper, we presented a proof-of-concept method for anomaly detection of smart grid control protocols. We observed inter-arrival times of IEC 104 communication and applied LOF algorithm for outlier detection. 
%to learn the packets inter-arrival time pattern and detect packets that fall out of the detected pattern. 
The preliminary results show that LOF creates a stable statistical model for ICS traffic and is able to detect outliers caused by cyber attacks. 
%We implemented the method as a proof-of-concept and used it in a set of experiments for demonstration purposes. 
The proposed security mechanism was tested and validated using our IEC 104 datasets. Experiments show the suitability, usability and high accuracy of the proposed method on smart grid communication. % Although only the IEC 104 protocol was covered in this research, 
In the future work we plan to apply this approach to other ICS protocols like GOOSE, MMS and DLMS. We also plan to observe other statistical features like packet or flow size. Although our key goal was to test the applicability of the proposed solution, we also tested the performance of the developed tool. We gathered basic time statistics. The processing time of extracting  key attributes from the PCAP file into a CSV file took an average of one minute and 10 seconds. This time had small deviations because it takes time to initialize Tshark. The time to learn a CSV file depends on the number of packets. At average, to learn 5,000 packets, 20 seconds were required. 
%In future research, we plan to accelerate the computation speed of our tool. 

The results are part of my PhD research that focused on statistical-based anomaly detection in ICS protocols. In the future work, we focus is extending detection model to  common cyber attacks defined by MITRE ATT\&CK ICS matrix. 

\section*{Acknowledgment}
The work is supported by the Brno University of Technology project "Application of AI methods to cyber security and control systems", no. FIT-S-20-6293.



\bibliographystyle{IEEEtran}
\bibliography{references.bib}

\begin{thebibliography}{10}
\providecommand{\url}[1]{#1}
\csname url@samestyle\endcsname
\providecommand{\newblock}{\relax}
\providecommand{\bibinfo}[2]{#2}
\providecommand{\BIBentrySTDinterwordspacing}{\spaceskip=0pt\relax}
\providecommand{\BIBentryALTinterwordstretchfactor}{4}
\providecommand{\BIBentryALTinterwordspacing}{\spaceskip=\fontdimen2\font plus
\BIBentryALTinterwordstretchfactor\fontdimen3\font minus
  \fontdimen4\font\relax}
\providecommand{\BIBforeignlanguage}[2]{{%
\expandafter\ifx\csname l@#1\endcsname\relax
\typeout{** WARNING: IEEEtran.bst: No hyphenation pattern has been}%
\typeout{** loaded for the language `#1'. Using the pattern for}%
\typeout{** the default language instead.}%
\else
\language=\csname l@#1\endcsname
\fi
#2}}
\providecommand{\BIBdecl}{\relax}
\BIBdecl

\bibitem{leith-13-identification}
H.~Leith and J.~W. Piper, ``{Identification and application of security
  measures for petrochemical industrial control systems},'' \emph{Journal of
  Loss Prevention in the Process Industries}, vol.~26, no.~6, pp. 982--993,
  2013.

\bibitem{lee-16-analysis}
R.~M. Lee, M.~J. Assante, and T.~Conway, ``{Analysis of the Cyber Attack on the
  Ukrainian Power Grid. Defense Use Case},'' Electricity Information Sharing
  and Analysis Center, Tech. Rep., March 2016.

\bibitem{tidy_2021}
\BIBentryALTinterwordspacing
J.~Tidy, ``Us fuel pipeline 'paid hackers \$5m in ransom','' 2021. [Online].
  Available: \url{https://www.bbc.com/news/business-57112371}
\BIBentrySTDinterwordspacing

\bibitem{matousek-19-increasing}
P.~Matou\v{s}ek, O.~Ry\v{s}av\'{y}, and M.~Gr\'{e}gr,
  ``\BIBforeignlanguage{english}{{Increasing Visibility of IEC 104
  Communication in the Smart Grid}},'' in
  \emph{\BIBforeignlanguage{english}{The 6th International Symposium for ICS \&
  SCADA Cyber Security Research 2019}}.\hskip 1em plus 0.5em minus 0.4em\relax
  BCS Learning and Development Ltd, 2019, pp. 21--30.

\bibitem{barbosa-12-first}
R.~R.~R. Barbosa, R.~Sadre, and A.~Pras, ``{A first look into SCADA network
  traffic},'' in \emph{2012 IEEE Network Operations and Management Symposium},
  April 2012, pp. 518--521.

\bibitem{barbosa-14-anomaly}
R.~R.~R. Barbosa, ``{Anomaly detection in SCADA systems: a network based
  approach},'' Ph.D. dissertation, University of Twente, 4 2014.

\bibitem{udd2016exploiting}
R.~Udd, M.~Asplund, S.~Nadjm-Tehrani, M.~Kazemtabrizi, and M.~Ekstedt,
  ``Exploiting bro for intrusion detection in a scada system,'' in
  \emph{Proceedings of the 2nd ACM International Workshop on Cyber-Physical
  System Security}, 2016, pp. 44--51.

\bibitem{bhatia2014practical}
S.~Bhatia, N.~S. Kush, C.~Djamaludin, A.~J. Akande, and E.~Foo, ``{Practical
  Modbus flooding attack and detection},'' in \emph{Proceedings of the 12th
  Australasian Information Security Conference}.\hskip 1em plus 0.5em minus
  0.4em\relax Australian Computer Society, Inc., 2014, pp. 57--65.

\bibitem{sayegh2014scada}
N.~Sayegh, I.~H. Elhajj, A.~Kayssi, and A.~Chehab, ``Scada intrusion detection
  system based on temporal behavior of frequent patterns,'' in \emph{MELECON
  2014-2014 17th IEEE Mediterranean Electrotechnical Conference}.\hskip 1em
  plus 0.5em minus 0.4em\relax IEEE, 2014, pp. 432--438.

\bibitem{barbosa2016exploiting}
R.~R.~R. Barbosa, R.~Sadre, and A.~Pras, ``Exploiting traffic periodicity in
  industrial control networks,'' \emph{International journal of critical
  infrastructure protection}, vol.~13, pp. 52--62, 2016.

\bibitem{yang2014stateful}
Y.~Yang, K.~McLaughlin, S.~Sezer, Y.~Yuan, and W.~Huang, ``Stateful intrusion
  detection for iec 60870-5-104 scada security,'' in \emph{2014 IEEE PES
  General Meeting}.\hskip 1em plus 0.5em minus 0.4em\relax IEEE, 2014, pp.
  1--5.

\bibitem{goldenberg2013accurate}
N.~Goldenberg and A.~Wool, ``Accurate modeling of modbus/tcp for intrusion
  detection in scada systems,'' \emph{international journal of critical
  infrastructure protection}, vol.~6, no.~2, pp. 63--75, 2013.

\bibitem{kleinmann2016automatic}
A.~Kleinmann and A.~Wool, ``Automatic construction of statechart-based anomaly
  detection models for multi-threaded scada via spectral analysis,'' in
  \emph{Proceedings of the 2nd ACM Workshop on Cyber-Physical Systems Security
  and Privacy}, 2016, pp. 1--12.

\bibitem{caselli2015sequence}
M.~Caselli, E.~Zambon, and F.~Kargl, ``Sequence-aware intrusion detection in
  industrial control systems,'' in \emph{Proceedings of the 1st ACM Workshop on
  Cyber-Physical System Security}, 2015, pp. 13--24.

\bibitem{knapp2014industrial}
E.~D. Knapp and J.~T. Langill, \emph{Industrial Network Security: Securing
  critical infrastructure networks for smart grid, SCADA, and other Industrial
  Control Systems}.\hskip 1em plus 0.5em minus 0.4em\relax Syngress, 2014.

\bibitem{FITPUB12385}
P.~Matou\v{s}ek, V.~Havlena, and L.~Hol\'{i}k,
  ``\BIBforeignlanguage{english}{Efficient modelling of ics communication for
  anomaly detection using probabilistic automata},'' in
  \emph{\BIBforeignlanguage{english}{IFIP/IEEE Int. Symposium on Integrated
  Network Management}}, 2021.

\bibitem{breunig-00-lof}
M.~M. Breunig, H.-P. Kriegel, R.~T. Ng, and J.~Sander, ``{LOF: Identifying
  Density-Based Local Outliers},'' \emph{SIGMOD Rec.}, vol.~29, no.~2, 2000.

\bibitem{boukerche2020outlier}
A.~Boukerche, L.~Zheng, and O.~Alfandi, ``Outlier detection: Methods, models,
  and classification,'' \emph{ACM Computing Surveys (CSUR)}, vol.~53, no.~3,
  pp. 1--37, 2020.

\bibitem{alghushairy-21-review}
O.~Alghushairy, R.~Alsini, T.~Soule, and X.~Ma, ``{A Review of Local Outlier
  Factor Algorithms for Outlier Detection in Big Data Streams},'' \emph{Big
  Data and Cognitive Computing}, vol.~5, no.~1, 2021.

\end{thebibliography}

\end{document}